\documentclass[12pt]{article}
\usepackage{latexsym}

\newcommand{\be}{\begin{equation}}
\newcommand{\ee}{\end{equation}}
\newcommand{\bea}{\begin{eqnarray}}
\newcommand{\eea}{\end{eqnarray}}

\newcommand{\etc}{{\it etc}}
\newcommand{\str}{{\rm str\;}}

\newcommand{\qbar}{{\overline{q}}}
\newcommand{\sbar}{{\overline{s}}}
\newcommand{\psibar}{{\overline{\psi}}}
\newcommand{\tq}{\tilde{q}}
\newcommand{\td}{\tilde{d}}
\newcommand{\tqbar}{{\overline{\tq}}}
\newcommand{\tdbar}{{\overline{\td}}}
\newcommand{\Nhat}{\hat{N}}
\newcommand{\tQ}{\tilde{Q}}
\newcommand{\tK}{\tilde{K}}

\begin{document}

\begin{titlepage}

%\begin{flushright} \\
%\today
%\end{flushright}
%\vspace*{1.5cm}
\begin{center}
{\Large \bf Analytic estimates for penguin operators \\
[.5cm]in quenched QCD}\\[3.0cm]

{\bf Maarten Golterman}$^{a,b}$ and {\bf Santiago Peris}$^b$\\[1cm]

$^a$Department of Physics and Astronomy, San Francisco State University%
\footnote{permanent address}\\
1600 Holloway Ave, San Francisco, CA 94132, USA\\
e-mail: {\tt maarten@quark.sfsu.edu}\\[0.5cm]
$^b$Grup de F{\'\i}sica Te{\`o}rica and IFAE\\ Universitat
Aut{\`o}noma
de Barcelona, 08193 Barcelona, Spain\\
e-mail: {\tt peris@ifae.es}

\end{center}

\vspace*{1.0cm}

\begin{abstract}

Strong penguin operators are singlets under the right-handed flavor
symmetry group SU(3)$_R$.  However, they do not remain singlets when
the operator is embedded in (partially)
quenched QCD, but instead they become linear
combinations of two operators with different transformation
properties under the (partially) quenched symmetry group.
This is an artifact of the quenched approximation.  Each of these
two operators is represented by a different set of low-energy constants
in the chiral effective theory.  In this paper, we give analytic
estimates for the leading low-energy constants, in quenched and
partially quenched QCD.  We conclude that the
effects of quenching on $Q_6$ are large.

\end{abstract}

\end{titlepage}

\noindent

Even at present, the number of light dynamical (or sea) quarks in
Lattice QCD computations is often not equal to that of the real world.  In
particular, in quite a few of the more difficult computations,
no sea quarks have been taken into account at all --- in other words, they
have been done in the quenched approximation.

The adaptation of chiral perturbation theory (ChPT) to the quenched
theory has been useful for gaining insight into the effects of quenching.
However, ChPT itself does not give any insight into the values of
the parameters of the effective theory, the so-called low-energy constants
(LECs).  It follows that ChPT also does not tell us anything about
the effects of quenching on the values of the LECs.  Within Lattice QCD,
therefore, we do in principle not know anything systematic about the errors
due to quenching, until computations are done with dynamical quarks,
and the results are compared with those of quenched QCD.

At the same time, analytic approaches are being explored to obtain
estimates for, in particular, electroweak-interaction LECs.  Some of these
approaches make sophisticated use of analytic knowledge available
about QCD, such as its chiral behavior, the operator-product expansion,
and large-$N_c$ techniques.   For the purpose of this paper, a relevant
reference is Ref.~\cite{hpdr}, to which we also refer for references
to other work.  While some assumptions have to be made
in such approaches because a non-perturbative analytic solution to
QCD is not available, they are often so tightly constrained that it
is quite likely that the results obtained from them will help us with
our quantitative understanding of hadron phenomenology.  Moreover,
they have the advantage of exhibiting the underlying reasons for the
size of specific contributions.

It is therefore natural to adapt these analytic techniques to
include the effects of (partial) quenching.  This is useful, because
it gives us {\it quantitative} information about the effects of quenching.
Results from this approach can thus serve as a guide to the merits
and pitfalls of the quenched approximation.

In this note, we apply the large-$N_c$ expansion to the calculation of the
leading LECs associated with the strong penguin $Q_6$
\cite{svz}.  It was
recently pointed out that even the definition of the quenched
version of this operator is ambiguous \cite{gppenguin}, and that
as a consequence, the quenched theory has more LECs associated with
this operator than the unquenched theory.  It turns out that the
leading-order LECs can be estimated analytically if one ignores
order $1/N_c^2$ corrections.  These estimates tell us that
the effects of quenching on the contribution of $Q_6$ to
non-leptonic kaon decays are likely to be large.

The operator $Q_6$ is defined as
\be
Q_6=4(\sbar_L^\alpha\gamma_\mu d_L^\beta)\sum_{q=u,d,s}
(\qbar_R^\beta\gamma_\mu q_R^\alpha)\,,
\label{q6}
\ee
where $q_{R,L}=P_{R,L}q$ with $P_{R,L}=
\frac{1}{2}(1\pm\gamma_5)$, and $\alpha$ and $\beta$
are color indices.  This operator transforms in the $(8,1)$ representation
of SU(3)$_L\times$SU(3)$_R$.

In order to ``embed" this operator in the quenched theory, let us
briefly recall how one may define quenched QCD as a field theory
\cite{morel,bgq}.   For each quark $q$ one introduces a ghost
quark $\tq$ with the same mass, spin, and color, but opposite
statistics.  The opposite statistics cause the path integral
over the ghost quarks to cancel the quark determinant (for each
gauge field configuration), effectively replacing the quark
determinant by one.  This is precisely the definition of the
quenched approximation.  It follows that the flavor symmetry
of the quenched theory is not described by SU(3), but by the
larger, graded group SU(3$|$3).  $Q_6$ is not a singlet under
SU(3$|$3)$_R$, but instead can be decomposed as \cite{gppenguin}
\bea
Q_6&=&\frac{1}{2}\;Q_6^{QS}+Q_6^{QNS}\,,
\label{decomp}\\
Q_6^{QS}&=&4(\sbar_L^\alpha\gamma_\mu d_L^\beta)\sum_\psi
(\psibar_R^\beta\gamma_\mu\psi_R^\alpha)\,,\nonumber\\
Q_6^{QNS}&=&4(\sbar_L^\alpha\gamma_\mu d_L^\beta)\sum_\psi
(\psibar_R^\beta\gamma_\mu\Nhat\psi_R^\alpha)\,,\nonumber\\
\Nhat&=&\frac{1}{2}\;{\rm diag}\;(1,1,1,-1,-1,-1)\,,\nonumber
\eea
where $\psi=(q,\tq)$.  $Q_6^{QS}$ clearly is a singlet under
SU(3$|$3)$_R$, but $Q_6^{QNS}$ is not.

To leading order in quenched ChPT, these operators are bosonized by
(we work in euclidean space)
\bea
Q_6^{QS}&\to& -\alpha^{(8,1)}_{q1}\str(\Lambda L_\mu L_\mu)
+\alpha^{(8,1)}_{q2}\str(2B_0\Lambda(\Sigma M+M\Sigma^\dagger))\,,
\label{bosonized}\\
Q_6^{QNS}&\to& f^2\alpha^{NS}_q\str(\Lambda\Sigma\Nhat\Sigma^\dagger)\,.
\nonumber
\eea
In these expressions $\Sigma=\exp(2i\Phi/f)$ is the non-linear field
describing the quenched Goldstone-meson multiplet, $M$ is the quark
mass matrix, $B_0$ is defined in Ref.~\cite{gl}, and
$f$ is the chiral limit of the pion decay constant normalized such
that $f_\pi=132$~MeV.  $\Lambda$ is the tensor picking out the
octet operator $\sbar d=\qbar\Lambda q$, and $L_\mu=i\Sigma\partial_\mu
\Sigma^\dagger$ is the left-handed current to leading order in ChPT.
$\alpha^{(8,1)}_{q1,2}$ are the LECs associated with the order $p^2$
weak octet kinetic and mass terms \cite{cbetal}, while $\alpha^{NS}_q$
is the order $p^0$ LEC associated with the non-singlet operator
$Q_6^{QNS}$ \cite{gppenguin}.  Note that, while this operator is
enhanced in ChPT relative to the singlet operator, its contributions
to matrix elements with only physical quarks on the external lines
start only at order $p^2$ \cite{gppenguin,
gppenguin2}.  The subscript $q$ is there as a reminder that the
$\alpha$'s refer to LECs of the quenched theory.

We begin with estimating the magnitude of $\alpha^{NS}_q$.  A simple
way of doing this was outlined in Ref.~\cite{gppenguin2}.  One first
rotates $Q_6^{QNS}$ by an SU(3$|$3)$_L$ rotation into
\bea
\tQ_6^{QNS}&=&4(\sbar_L^\alpha\gamma_\mu\td_L^\beta)
\sum_\psi(\psibar_R^\beta\gamma_\mu\Nhat\psi_R^\alpha)\,,
\label{q6tilde}\\
&=&-4\left((\sbar P_R q)(\qbar P_L\td)+(\sbar P_R\tq)(\tqbar P_L\td)\right)
\nonumber
\eea
where in the second line we fierzed the operator, paying careful
attention to the fact that the ghost-quark fields are commuting.
One then considers the $\tK^0\to 0$ matrix element of this operator,
with $\tK^0$ a hybrid kaon made of a ghost-$d$ quark and
a physical anti-$s$ quark.  The advantage of considering this
matrix element is that it is of order $p^0$ in ChPT \cite{gppenguin2}:
\be
\langle 0|\tQ_6^{QNS}|\tK^0\rangle=2if\alpha^{NS}_q+O(p^2)\,.
\label{chpt}
\ee

It turns out to be quite simple to find an expression for $\alpha^{NS}_q$.
By carrying out Wick contractions,
we may write the $\tK^0\to 0$ matrix element as
\bea
\langle 0|\tQ_6^{QNS}|\tK^0\rangle&=&
(\langle\sbar s\rangle
-\langle\tdbar\td\rangle)\langle 0|\sbar\gamma_5\td|\tK^0\rangle
\left(1+O\left(\frac{1}{N_c^2}\right)\right)\,\label{contr}\\
&&-4\langle 0|(\sbar P_R\,
\underline{q\qbar}\,P_L\td+\sbar P_R\,
\underline{\tq\tqbar}\,P_L\td)|\tK^0\rangle\,,
\nonumber
\eea
where $\underline{q\qbar}$ denotes the contraction of $q$ with
$\qbar$, and likewise for $\underline{\tq\tqbar}$.
The first line contains terms with two quark loops, while the
second term has only one quark
loop.  In order to connect the two loops on the first line, at least
two gluon lines are needed to obtain a non-zero contribution, making
such connected contributions suppressed by $1/N_c^2$.  To leading
order in $1/N_c$ we thus obtain the factorized contributions shown
explicitly.

The one-loop contribution maybe dealt with as follows.  First,
we observe that ghost and physical quark propagators are equal,
flavor by flavor, by construction.  Then, we may, in this term,
rotate the $\tK^0$ back to a $K^0$, and correspondingly the
$\td_L$ quark to $d_L$.  For the second term we find
\bea
-4\langle 0|(\sbar P_R\,
\underline{q\qbar}\,P_L\td+\sbar P_R\,
\underline{\tq\tqbar}\,P_L\td)|\tK^0\rangle
&=&-8\langle 0|\sbar P_R\,\underline{q\qbar}\,P_L d|K^0\rangle
\label{unf}\\
&=&-8\langle 0|(\sbar P_R q)(\qbar P_L d)|K^0\rangle\,.\nonumber
\eea
The last step follows because, as one can show, the Wick
contractions leading to contributions with two quark loops
cancel each other in the chiral limit.  The last expression
in Eq.~(\ref{unf}) is just the $K^0\to 0$ matrix element of $Q_6$,%
\footnote{through the weak mass term \cite{cbetal}.}
which is of order $p^2$.
This is also true in the quenched theory
\cite{gppenguin}.  Hence, this term does not contribute to $\alpha^{NS}_q$.

Using that, in the chiral limit, $\langle\tdbar\td\rangle=
-\langle\sbar s\rangle=\frac{1}{2}f^2B_0$,
we thus find for the matrix element to order $p^0$ that
\bea
\langle 0|\tQ_6^{QNS}|\tK^0\rangle&=&
2\langle\sbar s\rangle\langle0|\sbar\gamma_5\td|\tK^0\rangle
\label{fact}\\
&=&2\langle\sbar s\rangle\; if\frac{M_K^2}{m_s+m_d}=-if^3B_0^2\,.
\nonumber
\eea
Comparing with Eq.~(\ref{chpt}),
we obtain an estimate for $\alpha^{NS}_q$ correct to order
$1/N_c^2$:
\be
\alpha^{NS}_q=-\frac{1}{2}f^2B_0^2\,.
\label{alphans}
\ee

In order to get an idea about the value of $\alpha^{NS}_q$, we may compare it to the
value of $\alpha^{(8,1)}_{q1}$. Fortunately, it turns out to be remarkably simple to
obtain an estimate for $\alpha^{(8,1)}_{q1}$ in the quenched theory. It turns out that in
the quenched case, the unfactorized contribution vanishes! This can be seen as follows.
The fierzed form of the singlet operator $Q_6^{QS}$ is, from Eq.~(\ref{decomp}), \be
Q_6^{QS}=-8\left((\sbar P_R q)(\qbar P_L d) -(\sbar P_R \tq)(\tqbar P_L d)\right)
\,.\label{fsinglet} \ee Following the analysis of Ref.~\cite{hpdr},\footnote{As in
Ref.~\cite{hpdr}, we will work in the leading-log approximation, in which $Q_6^{QS}$ does
not mix with any other operator in the quenched theory.} again there is a contribution
with two quark loops, and a contribution with one quark loop. The one-loop contribution
again corresponds to the terms in which $q_R$ and $\qbar_R$ or $\tq_R$ and $\tqbar_R$ are
contracted, and thus vanishes because of the relative minus sign in Eq.~(\ref{fsinglet}).
This leaves us with the two-loop contribution, which factorizes to order $1/N_c^2$,
yielding \cite{factorized} \be \alpha^{(8,1)}_{q1}=-8L_5 f^2 B_0^2 \label{alpha8} \ee
(equivalent to $g_8=-16L_5 B_0^2/F_0^2$, $F_0=f/\sqrt{2}$
in the notation of Ref.~\cite{hpdr}).%
\footnote{$L_5$ is one of the Gasser--Leutwyler constants \cite{gl}.} Note that quenched
$L_5$ does not run.

Putting things together, we find that \be
\frac{\alpha^{NS}_q}{\alpha^{(8,1)}_{q1}}=\frac{1}{16L_5}
\left(1+O\left(\frac{1}{N_c^2}\right)\right)\,. \label{ratio} \ee Our first conclusion
based on these results is that $\alpha^{NS}_q$ is likely to be large compared to the
singlet LECs. The value of $L_5$ is of order $10^{-3}$ both in the quenched \cite{l5q}
and unquenched \cite{gl} theories, making this ratio of order 60.  This is not
small,\footnote{In particular, ``strategy 3" of Ref.~\cite{gppenguin2} would yield
results significantly different from the other strategies.} and casts doubt on the
tentative conclusion of  Ref.~\cite{rbc} (concluding section) on the size of
$\alpha^{NS}_q$. If $\alpha^{NS}_q$ is not small, this could have a dramatic effect on
the extraction of $\alpha^{(8,1)}_{q2}$ from $K^0\to 0$, which in turn is needed for the
extraction of $\alpha^{(8,1)}_{q1}$ from $K\to\pi$ \cite{cbetal}, because it appears at
the same order as $\alpha^{(8,1)}_{q2}$ in ChPT in the $K^0\to 0$ matrix element
\cite{gppenguin}.

Another interesting observation is that the value of $\alpha^{(8,1)}_{q1}$
maybe significantly smaller (in absolute value) than that of the unquenched
theory.  This is because of the absence of unfactorized contributions
in the quenched theory, which,
in the unquenched theory, have been estimated to be sizeable compared to the
factorized contribution,
and of the same sign \cite{hpdr}.\footnote{The unfactorized
contribution is found to be about twice the factorized one, at the
scale of the $\rho$ mass.}  Even if one would decide that one should define $Q_6$ in the quenched theory to be $Q_6^{QS}$ only,
omitting $Q_6^{QNS}$ altogether,
as an alternative possibility \cite{gppenguin,gppenguin2},%
\footnote{This change is known to have a large effect
\cite{lanl,cppacs,gpproc}.}
this smaller value of $\alpha^{(8,1)}_{q1}$ would lead to a reduction
of the value of $\varepsilon'/\varepsilon$ in quenched QCD,
relative to its unquenched value.

It is rather easy to extend these estimates to the partially
quenched situation, in which $N$ (massless) sea quarks are added
to the quenched theory \cite{bgpq}.  First, the non-singlet part of
$Q_6$ is now represented by the order $p^0$ LEC $\alpha^{(8,8)}$
\cite{gppenguin},
because in the partially quenched case the non-singlet operator is
in the same irreducible representation as $Q_8$ \cite{gppenguin}
(and thus it corresponds to
$g_{\rm ew}$ of Ref.~\cite{hpdr}).
One finds that the same expression as given for $\alpha^{NS}_q$
in Eq.~(\ref{alphans}) is also valid in the partially quenched theory
for $\alpha^{(8,8)}$.
For $\alpha^{(8,1)}_1$, a naive estimate would be to interpolate linearly
in the number of sea-quark flavors $N$ between the quenched and
unquenched theories.  Using the results obtained in Ref.~\cite{hpdr},
which considers the unfactorized contribution proportional to the number
of light flavors (which is three in the real world), this would lead to the
estimate
\be
\alpha^{(8,1)}_1=-8L_5 f^2 B_0^2+\frac{N}{3}\times
{\rm (unfactorized\ contribution\ of\ Ref. ~\cite{hpdr})}\,,
\label{alphapq}
\ee
where of course now $L_5$, $f$, $B_0$, \etc. take their values in the
partially quenched theory with $N$ sea quarks.

Finally, let us comment on the fact that, since our aim was to extract values
for the leading order LEC for each of the operators we considered,
all calculations were done in the chiral limit.
It is well known that the chiral limit of the quenched theory is
hampered by severe infrared divergences, and, in fact, the
chiral limit of the quenched theory may not exist \cite{bgq,sharpe,bgqhs}.
However, we believe that our results for LECs are not affected by this issue.
If the quenched effective theory makes any sense, its parameters,
which are the LECs, should be well defined and finite in the chiral
limit.  All the sicknesses associated with the quenched infrared behavior
should be correctly reproduced if the appropriate fields,
in particular the $\eta'$, are kept in the effective theory.

\vskip 1 cm

\leftline{\bf Acknowledgments}

\vspace*{3mm}
MG was supported in part by the US Dept.
of Energy, and SP was supported by
CICYT-FEDER-FPA2002-00748, 2001 SGR00188 and by TMR EC-Contracts
HPRN-CT-2002-00311 (EURIDICE).

\end{document}